\documentclass[]{raa}           

\input{epsf.sty}                        
\input{psfig.sty}                       
\usepackage{graphicx}
\usepackage{times}
\usepackage{float}
\usepackage{cmap}
\usepackage[T1]{fontenc}

\newcommand\nature{{Nature}}
\newcommand{\HII}{\ion{H}{ii}}

\newcommand{\OII}{[\ion{O}{iii}]}

\begin{document}

   \title{Kinematic properties of the dual AGN system J0038$+$4128 based on long-slit spectroscopy}

   \volnopage{Vol.0 (200x) No.0, 000--000}      
   \setcounter{page}{1}          

   \author{Yang-Wei Zhang
      \inst{1,2}
   \and Yang Huang
      \inst{3}
   \and Jin-Ming Bai
      \inst{1,4}
   \and Xiao-Wei Liu
    \inst{3,5}
    \and Jian-Guo Wang
    \inst{1,4}
   }

   \institute{Yunnan Observatories, Chinese Academy of Sciences, kunming, Yunnan 650011, China; {\it zhangyangwei@ynao.ac.cn; baijinming@ynao.ac.cn}\\
 \and
             University of Chinese Academy of Sciences,Beijing 100049, China\\
        \and
             Department of Astronomy, Peking University, Beijing 100871,China; {\it yanghuang@pku.edu.cn}\\
                     \and
             Key Laboratory for the Structure and Evolution of Celestial Objects, Chinese Academy of Sciences, Kunming 650011,China;\\
        \and
             Kavli Institute for Astronomy and Astrophysics, Peking University, Beijing 100871, China\\}

   \abstract{The study of kiloparsec-scale dual active galactic nuclei (AGN) will provide important clues to understand the co-evolution between the host galaxies and their central supermassive black holes undergoing a merging process.
We present long-slit spectroscopy of the J0038$+$4128, a kiloparsec-scale dual AGN candidate discovered by Huang et al. recently, using the Yunnan Faint Object Spectrograph and Camera (YFOSC) mounted on Li-Jiang 2.4-m telescope at Yunnan observatories.
From the long-slit spectra, we find that the average relative line-of-sight (LOS) velocity between the two nuclei (J0038$+$4128N and J0038$+$4128S) is about 150 km\,s$^{-1}$.
The LOS velocities of the emission lines from the gas ionized by the nuclei activities and of the absorption lines from stars governed by the host galaxies for different regions of the J0038$+$4128 exhibit the same trend.
The same velocities trend indicates that the gaseous disks are co-rotating with the stellar disks in this ongoing merge system.
We also find several knots/giant \HII\, regions scattered around the two nuclei with strong star formation revealed by the observed line ratios from the spectra.
Those regions are also detected clearly in {\it HST} $F336W/U$-band and {\it HST} $F555W/V$-band images.
\keywords{techniques: spectroscopic---
Galaxies: kinematics and dynamics---
galaxies: active ---
galaxies: interactions ---
galaxies: nuclei ---
galaxies: individual: J0038+4128 }}


   \maketitle



   \maketitle
\newpage
%
%
\section{Introduction}           
\label{sect:intro}

Almost all massive galaxies are believed to host a central super massive black holes (SMBHs) (Richstone et al. 1998). During the merger of gas-rich galaxies, active galactic nuclei (AGN) will be triggered because a large amount of gas could be sent to the central SMBHs by tidal interactions (Hernquist 1989; Kauffmann \& Haehnelt 2000; Hopkins et al. 2008). The dual AGN are therefore the natural products of two merging SMBHs triggered simultaneously by accreting gas in a gas-rich major merger (Begelman, Blandford
\& Rees 1980; Milosavljevi\'c \& Merritt 2001).
The search for dual AGN, especially the kiloparsec-scale ones, are of extreme importance to understand the relation between galaxy evolution and nuclei activities (Yu et al. 2011).
However, at present, no more than few dozens of dual AGN are found with separation between 1 and 10 kpc (e.g. Junkkarinen et al. 2001; Komossa et al. 2003; Ballo et al. 2004; Bianchi et al. 2008; Comerford et al. 2009a,b, 2011; Liu  et al. 2010, 2013; Fu et al. 2011a; Koss et al. 2011; Rosario et al. 2011; McGurk et al. 2011, 2012, 2014; Barrows et al. 2012; Huang et al. 2014, hereafter H14).
Moreover, the current studies are limited to identify the kiloparsec-scale dual AGN rather than investigate the co-evolution between host galaxies and central AGN.

Ionized gas and stellar kinematics of galaxies were used to check whether the merger processes is important in galaxy formation and evolution (e.g. Beltran et al. 2001).
Very recently, Vilforth \& Hamann (2015) investigated the ionized gas and stellar kinematics of four  double-peaked \OII$\lambda5007$ AGN.
They found that the ionized gas generally follows the stars except one show opposite trend between gas and stars.
The only system with misaligned kinematics of gas and stars had confirmed as kiloparsec-scale dual AGN using $Chandra$ X-ray data (Comerford et al. 2011).
This system also shows no obvious tidal feature that may indicate the timescale of its coalescence of binary black holes is longer than $\sim$\,1\,Gyr and the host galaxies are well relaxed.
Other three systems with same velocity trend between ionized gas and stars all show significant tidal features and are identified as dual AGN using optical or infrared data (McGurk et al. 2011; Fu et. al. 2011a; Fu et. al. 2012).
It is still difficulty to conclude that the aligned/misaligned kinematics of gas and stars is a general behavior of  early/late stages of dual AGN or just accidental.
Hence, more long-slit observations of confirmed dual AGN will be great helpful to test this scenario.

\begin{figure}
  \centering
  \includegraphics[width=7cm,height=9cm]{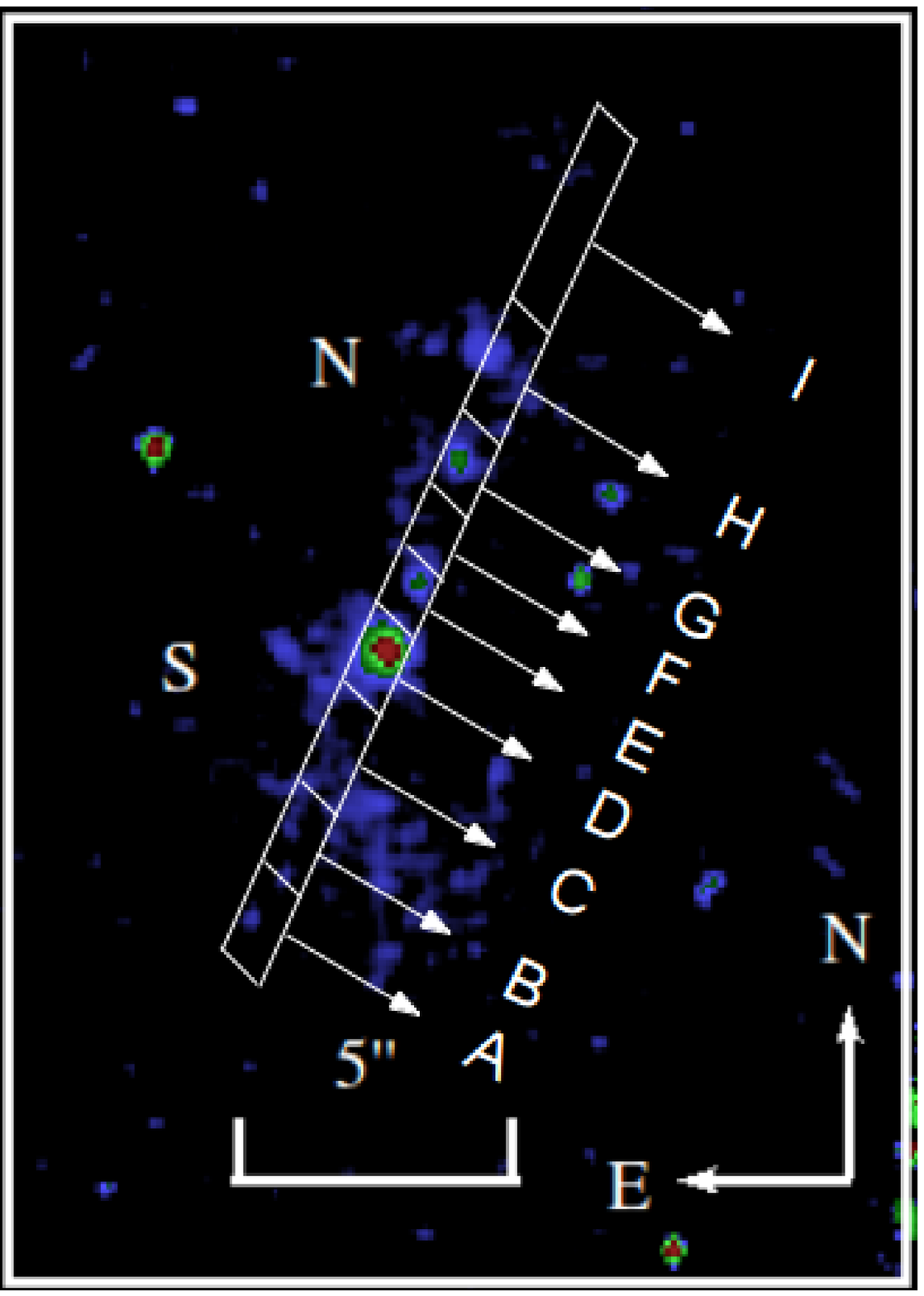}
  \includegraphics[width=7cm,height=9cm]{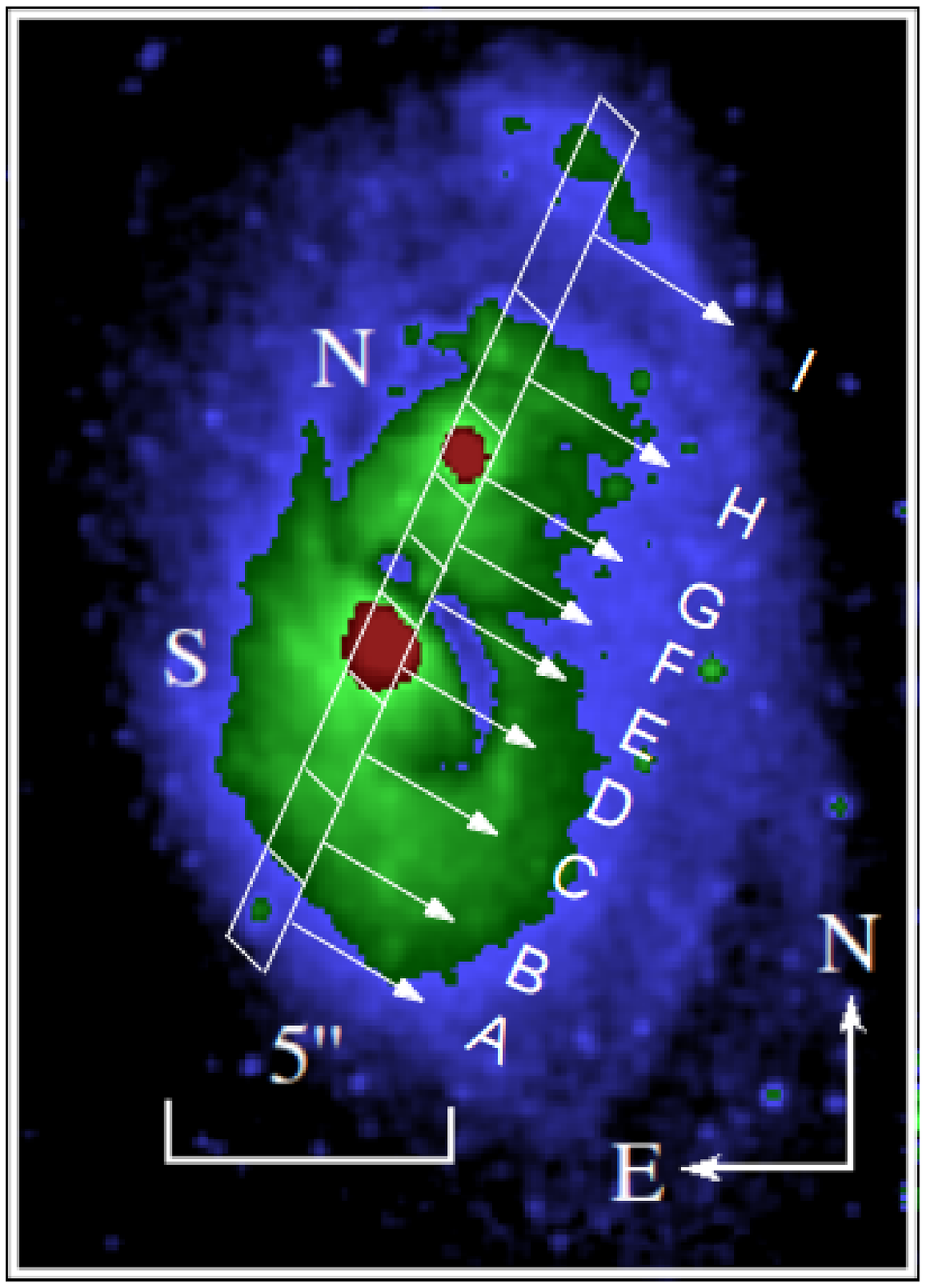}
\caption{The images of J0038+4128 are similar to the  Fig.\,\ref{slitimages} of H14 with {\it HST\/}/WFPC2 $F$336$W$/{\it U\/}-band pseudo colour image in left and  {\it HST\/}/WFPC2 $F$555$W$/{\it V\/}-band pseudo colour image in right. North is up and east is to the left. Spatial scale is also shown in each panel.
The slit position is set to cross the two nuclei.
The slit is further  divided into nine slices labeled as different letters to extract 1D spectra of interesting regions.}
\label{slitimages}
\end{figure}

Here, we present detailed kinematic analysis based on long-slit spectroscopy of J0038+4128, which is a kiloparsec-scale dual AGN candidate discovered by H14 recently.
J0038+4128 ($z$ = 0.0725) was confirmed as a Seyfert 1--Seyfert 2 dual AGN with two clear optical
nuclei from the Hubble Space Telescope ({\it HST}) Wide Field Planetary
Camera 2 (WFPC2) images (see Fig.\, \ref{slitimages}) of a small projected separation of 4.7 kpc.
The southern component (J0038+4128S) is confirmed as a Seyfert\,1
galaxy with broad Ly$\alpha$ emission line (H14). The northern component (J0038+4128N) is confirmed as a Seyfert\,2 galaxy with narrow lines (H14).
This system is suffering strong interactions and shows two pairs of
bi-symmetric arms  observed in dual AGN for the first time.
J0038+4128 is very bright and thus allows us to obtain high quality spectra to study the kinematics of both ionized gas and stars, which also can enable us to investigate the relations amongst star formation, nuclear activity and galaxies evolution (Keel, William C. 1996; Colpi \& Dotti 2011; Yu et al. 2011; Shields et al. 2012; Kormendy \& Ho 2013).

The paper is organized as follows: The observations and data reductions are described in Section 2. The results, as well as discussions, are presented in Section 3. Finally, we summarize our main conclusions in Section 4.
We adopt a $\Lambda$CDM cosmology with $\Omega_{\rm m}$ = 0.3, $\Omega_{\Lambda}$ = 0.7, and $H_0$ = 70 km s$^{-1} $Mpc$^{-1}$  throughout.
All quoted wavelengths are in air units.

\section{Observations \& Data Reductions}
\label{sect:Obs and data}
Data were taken on the night of 2013 November 10 at the  2.4-m  telescope on Li-Jiang observatory (hereafter LJT; Zhang et al. 2012).
The weather of that night was clear and of excellent seeing ($\sim$\,1\farcs0).
To fully subtract the cosmic rays and achieve high spectral signal-to-noise (S/N) ratio, three exposures were taken with Yunnan Faint Object Spectrograph and Camera (YFOSC; Zhang et al. 2012) G8 grating through a slit width of 1\farcs0 ($R \sim 2200$, $\lambda = 5100-9600{\rm \AA}$) for a total integration time of 6900\,s (2100\,s, 2100\,s and 2700\,s for each exposure).
The pixel size of the YFOSC CCD chip is 0\farcs283 per pixel.
The slit position is shown in Fig.\ref{slitimages}, which is set to cross the centers of the two nuclei: J0038+4128N and J0038+4128S.

\begin{figure}
  \centering
  \includegraphics[width=12.6cm,height=6.0cm]{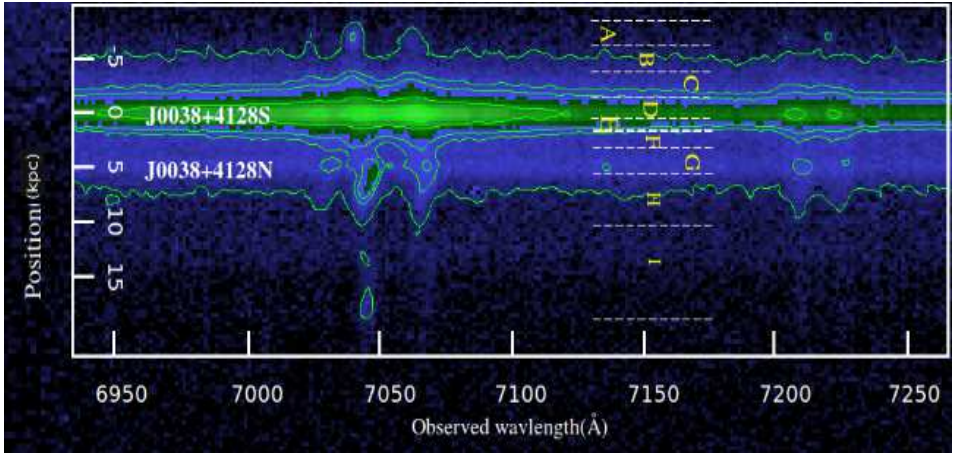}
\caption{Segment of the two-dimensional long-slit spectrum of J0038+4128 that exhibits spatially resolved  broad H${\alpha}$ emission line and narrow emission lines (H${\alpha}$+[N~{\sc ii}]+[S~{\sc ii}]) in the location of J0038+4128S and narrow emission lines (H${\alpha}$+[N~{\sc ii}]+[S~{\sc ii}]) in the location of J0038+4128N. The location of the different slices within the galaxy are labelled with letters and are shown in Fig. \ref{slitimages}.}
\label{slitspectra}
\end{figure}

All the data reduction were performed with IRAF and IDL.
The data were bias subtracted, flat-fielded calibrated, cosmic rays removed, wavelength calibrated and flux calibrated with ESO spectral flux standard star (BD+25 4655) also observed in that night.
The wavelength and flux calibrations were all performed in 2D.
The distortions of wavelength solution in spatial direction are also corrected with arc lines.
The final accuracies of the 2D wavelength calibrations are better than 5--10 km\,s$^{-1}$ checked by sky emission lines.
In order to better study the kinematics of J0038+4128, we extract 1D spectra of nine slices in spatial direction from the final 2D spectrum.
As shown in Figs.\,\ref{slitimages} \& \ref{slitspectra}, those slices represent different interesting regions (e.g. two nuclei, knots) of J0038+4128.
Here, we note that each slice contains spectra from at least 5 pixels and no point spread function (PSF) corrections are applied to the 1D spectra with a small seeing ($\sim$\, 1\farcs0).

\begin{figure}
  \centering
  \includegraphics[width=15cm,height=22cm]{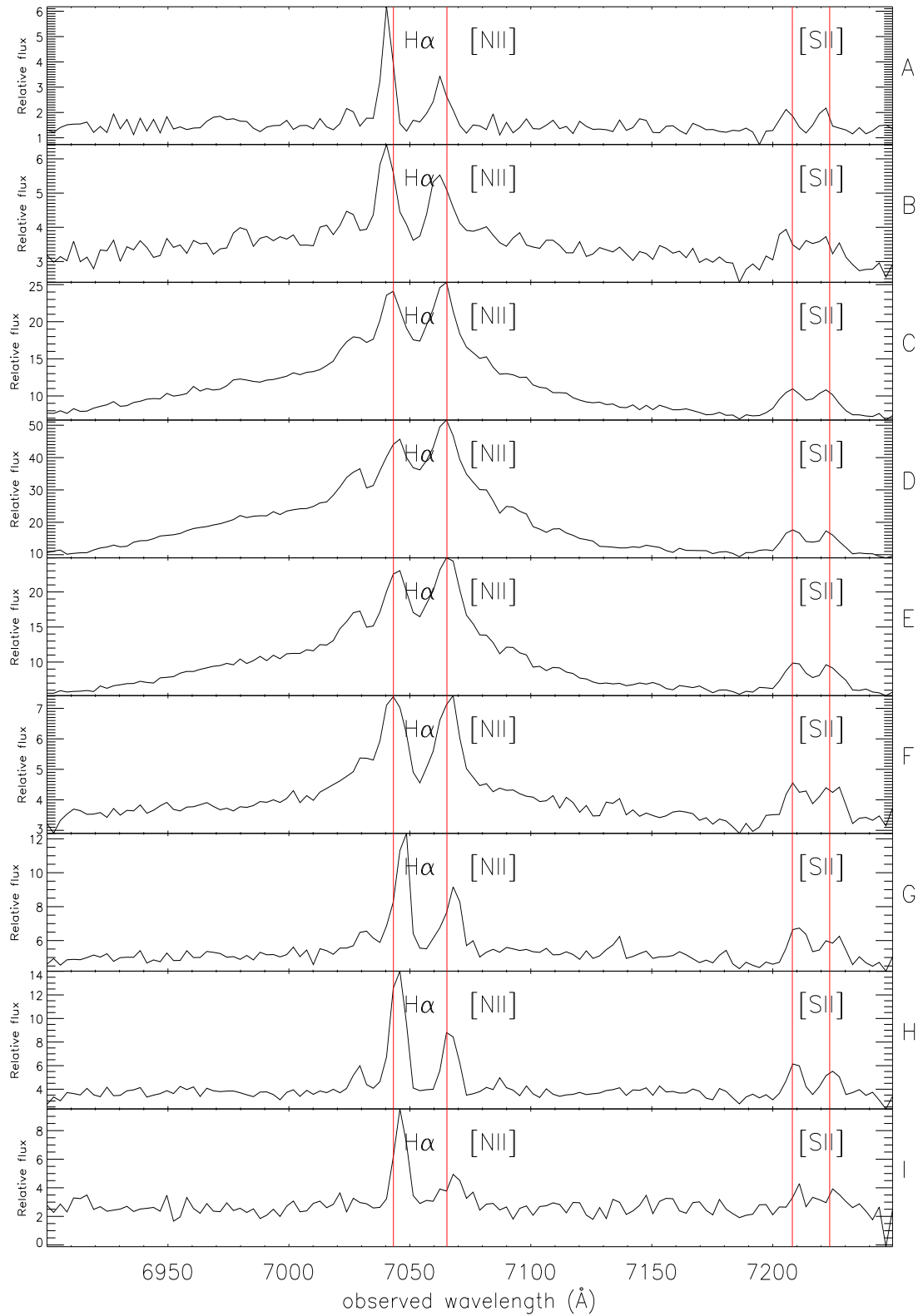}
  \caption{1D spectra extracted along the nine slices for J0038+4128 showing H${\alpha}$, $\left[ \textrm{NII} \right]$ and $\left[ \textrm{SII} \right]$ emission lines.
 Red lines indicate the detectable emission lines.
 Arbitrary flux units are shown for each panel to show relative line strengths.}

\label{line_p1}
\end{figure}

\begin{figure}
  \centering
  \includegraphics[width = 15 cm,height = 22 cm]{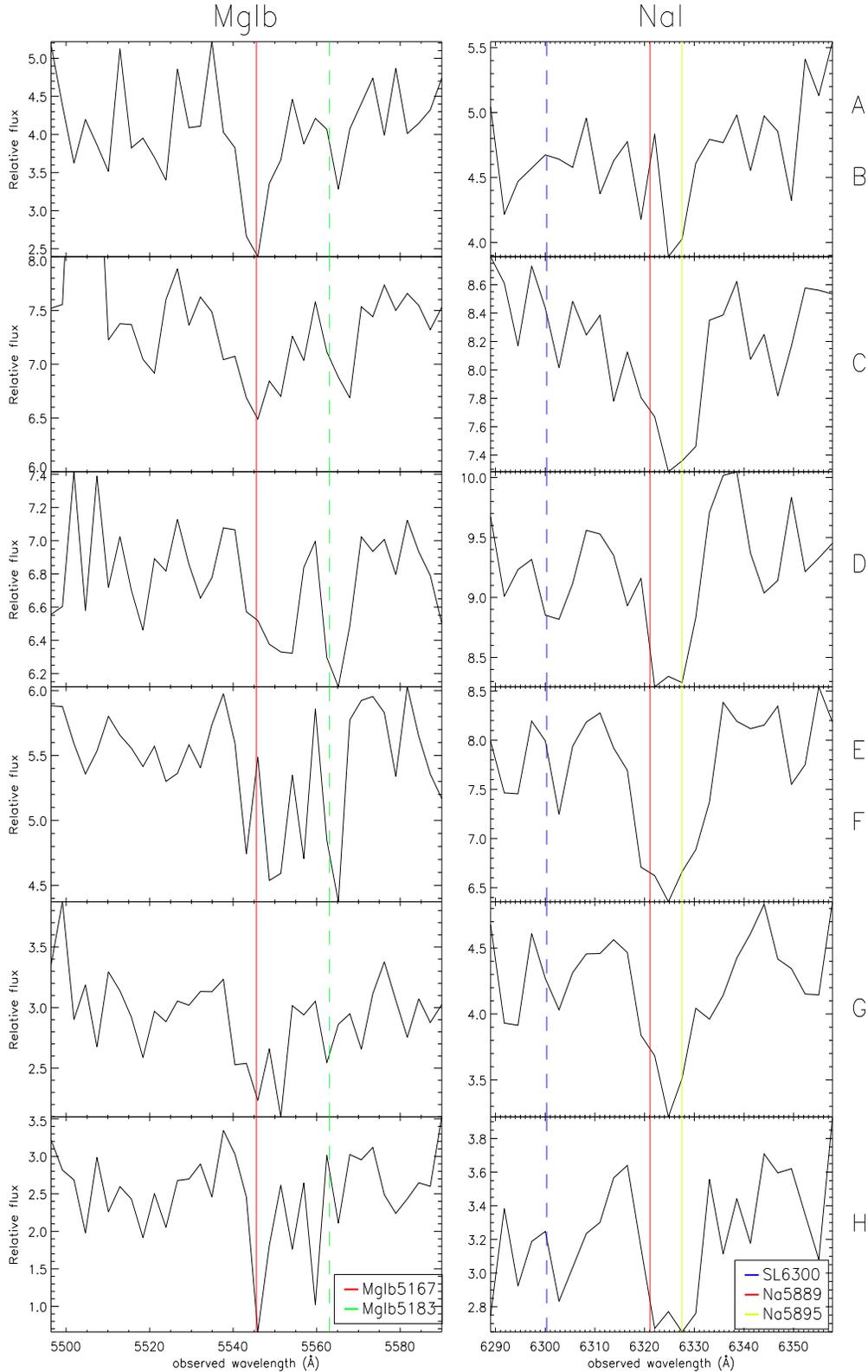}
\caption{1D spectra extracted along the nine slices for J0038+4128 showing MgIb$\lambda\lambda5167,5183$, NaI$\lambda\lambda5889,5895$ absorption lines.
Colourful lines indicate the existing absorption  lines.
 Arbitrary flux units are shown for each panel to show relative line strengths. SL6300 is the sky line in 6300 {\rm \AA}.}
\label{line_p2}
\end{figure}

\begin{table}[!htp]
\begin{center}
\caption{Spatially resolved kinematics of emission lines for J0038+4128. }
\begin{tabular}{lccccc}
\hline
Line & Slice & $V$ & $\Delta$ $V$ & FWHM& $\Delta$ FWHM \\
     &       &  (km s$^{-1}$) &  (km s$^{-1}$)  & (km s$^{-1}$) &  (km s$^{-1}$)  \\
\hline
H${\alpha}$    	&	   A   	&	-124 	&	3 	&	226 	&	11 	\\
H${\alpha}$    	&	   B   	&	-140 	&	7 	&	315 	&	17 	\\
H${\alpha}$    	&	   C   	&	-42 	&	11 	&	486 	&	28 	\\
H${\alpha}$    	&	   D   	&	26 	&	17 	&	486 	&	44 	\\
H${\alpha}$    	&	   E   	&	53 	&	10 	&	479 	&	27 	\\
H${\alpha}$    	&	   F   	&	9 	&	12 	&	444 	&	28 	\\
H${\alpha}$    	&	   G   	&	189 	&	4 	&	325 	&	13 	\\
H${\alpha}$    	&	   H   	&	88 	&	5 	&	317 	&	12 	\\
H${\alpha}$    	&	   I   	&	135 	&	3 	&	273 	&	10 	\\
H${\alpha}^{\rm broad}$  	&	   B   	&	-140 	&	7 	&	5841 	&	413 	\\
H${\alpha}^{\rm broad}$  	&	   C   	&	-42 	&	11 	&	6059 	&	218 	\\
H${\alpha}^{\rm broad}$  	&	   D   	&	26 	&	17 	&	5797 	&	233 	\\
H${\alpha}^{\rm broad}$  	&	   E   	&	53 	&	10 	&	5719 	&	198 	\\
H${\alpha}^{\rm broad}$  	&	   F   	&	9 	&	12 	&	5110 	&	309 	\\
$\left[ \textrm{NII} \right]$ 6583  &	   A   	&	-109 	&	9 	&	359 	&	41 	\\
$\left[ \textrm{NII} \right]$ 6583  &	   B   	&	-106 	&	12 	&	422 	&	28 	\\
$\left[ \textrm{NII} \right]$ 6583  &	   C   	&	-66 	&	10 	&	602 	&	26 	\\
$\left[ \textrm{NII} \right]$ 6583  &	   D   	&	-26 	&	16 	&	722 	&	42 	\\
$\left[ \textrm{NII} \right]$ 6583  &	   E   	&	12 	&	10 	&	643 	&	26 	\\
$\left[ \textrm{NII} \right]$ 6583  &	   F   	&	41 	&	9 	&	472 	&	21 	\\
$\left[ \textrm{NII} \right]$ 6583  &	   G   	&	111 	&	10 	&	469 	&	22 	\\
$\left[ \textrm{NII} \right]$ 6583  &	   H   	&	62 	&	10 	&	306 	&	22 	\\
$\left[ \textrm{NII} \right]$ 6583  &	   I   	&	125 	&	10 	&	418 	&	22 	\\
$\left[ \textrm{SII} \right]$ 6717 	&	    A   	&	-82 	&	13 	&	310 	&	30 	\\
$\left[ \textrm{SII} \right]$ 6717 	&	    B   	&	-156 	&	15 	&	272 	&	34 	\\
$\left[ \textrm{SII} \right]$ 6717 	&	    C   	&	-12 	&	12 	&	434 	&	27 	\\
$\left[ \textrm{SII} \right]$ 6717 	&	    D   	&	11 	&	17 	&	480 	&	41 	\\
$\left[ \textrm{SII} \right]$ 6717 	&	    E   	&	52 	&	13 	&	440 	&	30 	\\
$\left[ \textrm{SII} \right]$ 6717 	&	    F   	&	85 	&	22 	&	434 	&	52 	\\
$\left[ \textrm{SII} \right]$ 6717 	&	    G   	&	123 	&	13 	&	349 	&	30 	\\
$\left[ \textrm{SII} \right]$ 6717 	&	    H   	&	65 	&	14 	&	281 	&	32 	\\
$\left[ \textrm{SII} \right]$ 6717 	&	    I   	&	107 	&	10 	&	181 	&	23 	\\
$\left[ \textrm{SII} \right]$ 6731 	&	    A   	&	-129 	&	12 	&	252 	&	26 	\\
$\left[ \textrm{SII} \right]$ 6731 	&	    B   	&	-273 	&	28 	&	493 	&	70 	\\
$\left[ \textrm{SII} \right]$ 6731 	&	    C   	&	-87 	&	14 	&	484 	&	32 	\\
$\left[ \textrm{SII} \right]$ 6731 	&	    D   	&	-15 	&	18 	&	431 	&	43 	\\
$\left[ \textrm{SII} \right]$ 6731 	&	    E   	&	5 	&	15 	&	463 	&	37 	\\
$\left[ \textrm{SII} \right]$ 6731 	&	    F   	&	32 	&	25 	&	450 	&	60 	\\
$\left[ \textrm{SII} \right]$ 6731 	&	    G   	&	76 	&	23 	&	399 	&	55 	\\
$\left[ \textrm{SII} \right]$ 6731 	&	    H   	&	42 	&	23 	&	337 	&	53 	\\
$\left[ \textrm{SII} \right]$ 6731 	&	    I   	&	113 	&	18 	&	289 	&	41 	\\

\hline
\end{tabular}
\label{line_list}
\end{center}

 \begin{flushleft}
\quad \quad \quad \quad \quad \quad \quad \quad \quad
Notes: $V$: The LOS velocities are given with respect to average redshift ($z$ = 0.0732). \\ \quad \quad \quad \quad \quad \quad \quad \quad \quad $\Delta$ $V$: The errors of LOS velocities.\\
\quad \quad \quad \quad \quad \quad \quad \quad \quad FWHM: Full width at half-maximum.\\
\quad \quad \quad \quad \quad \quad \quad \quad \quad $\Delta$ FWHM: The errors of FWHM. 

 \end{flushleft}
\end{table}

\begin{table}[!htb]
\begin{center}
\caption{Spatially resolved kinematics of absorption lines for J0038+4128.}
\begin{tabular}{lccccc}
\hline
Line& Slice & $V$            & $\Delta$ $V$ &SNR\\
    &       &(km s$^{-1}$) &  (km s$^{-1}$)& ($\sigma$) \\
\hline

MgIb 5167 	&	 A+B   	&	-17 &	32  & 4.5	\\		
MgIb 5167 	&	 C   	&	7	&	29  & 3.1	\\	
MgIb 5167 	&	 D   	&	235 &	24  & 3.2	\\		
MgIb 5167 	&	 E+F   	&	241 &	22  & 4.1	\\		
MgIb 5167 	&	 G   	&	123 &	19  & 3.9	\\		
MgIb 5167 	&	 H   	&	59  &	25	& 3.8	\\	
MgIb 5167 	&	 I   	&	--  &	--  & --	\\		
\hline								
Na I 5889 	&	 A+B  	&	-98 &	27	& 4.2	\\	
Na I 5889 	&	 C   	&	-19 &	25	& 4.0	\\	
Na I 5889 	&	 D   	&	8	&	22	& 4.3	\\
Na I 5889 	&	 E+F   	&	77	&	28	& 8.4	\\
Na I 5889 	&	 G   	&	52	&	26	& 8.2	\\
Na I 5889 	&	 H   	&	36	&	22	& 4.7	\\
Na I 5889 	&	 I   	&	-- 	&	-- 	& --	\\
\hline								
Na I 5895 	&	 A+B   	&	-67 &	28	& 4.2	\\	
Na I 5895 	&	 C   	&	-45	&	24	& 4.0	\\
Na I 5895 	&	 D   	&	37	&	21	& 4.3	\\
Na I 5895 	&	 E+F   	&	52	&	23	& 8.4	\\
Na I 5895 	&	 G   	&	18	&	25	& 8.2	\\
Na I 5895 	&	 H   	&	-10	&	24	& 4.7	\\
Na I 5895 	&	 I   	&	-- 	&	-- 	& --	\\

\hline
\end{tabular}
\label{line_listb}
\end{center}

 \begin{flushleft}
\quad \quad \quad \quad \quad \quad \quad \quad \quad \quad Notes: $V$: The LOS velocities are given with respect to average redshift ($z$ = 0.0732). \\
\quad \quad \quad \quad \quad \quad \quad \quad \quad \quad  $\Delta$ $V$: The errors of LOS velocities.\\
\quad \quad \quad \quad \quad \quad \quad \quad \quad \quad  SNR: Signal-to-noise ratio.

 \end{flushleft}

\end{table}

\begin{figure}
  \centering
  \includegraphics[width = 15cm,height = 11cm]{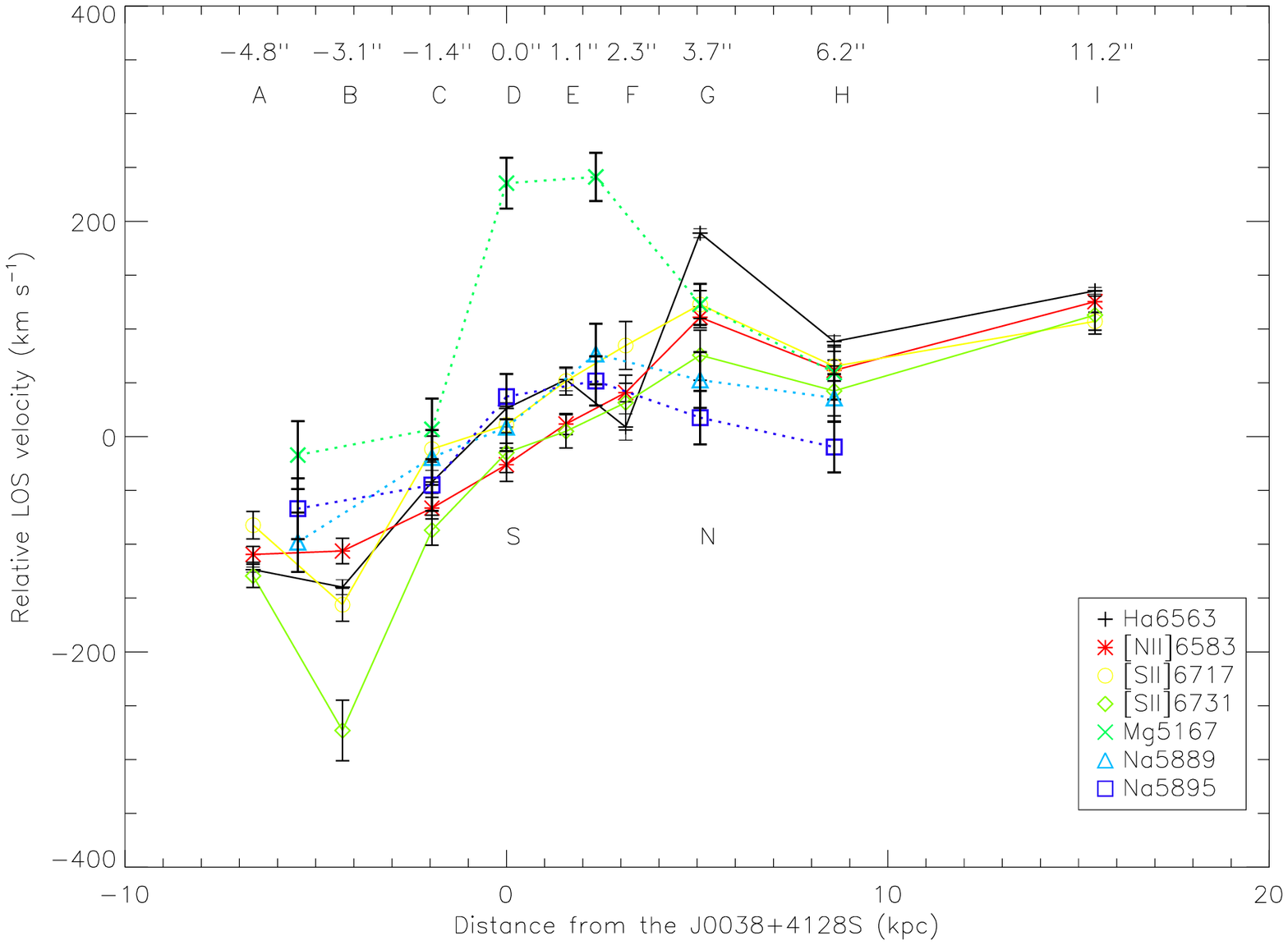}
\caption{Kinematic structure of different lines in J0038+4128, the location of the different slices within the galaxy are labelled with letters and are shown in Figs.\, \ref{slitimages}\,\& \ref{slitspectra}. N and S represent the cores of J0038+4128N and J0038+4128S respectively. The velocities are calculated with those emission lines respect to the reference redshift. For details about the parameters used in the line fits, see in Tables \,\ref{line_list}\& \ref{line_listb}.}

\label{velocity}
\end{figure}

\section{Results \& Discussions}
\label{Result_discussions}
\subsection{Kinematics}

As shown in Fig. \ref{slitspectra}, the 2D long-slit spectrum clearly exhibit two sets of spectra and rotation systems, which has already been discussed by H14 detailedly.
In the following subsections, we will focus on the kinematic properties of this system.
In principle, the [\ion{O}{iii}]$\lambda5007$ and H${\beta}$, emission lines are very important to study the kinematics of this system and are also included in our observed spectrum.
However, the instrumental efficiency  in the bluest range (5100\,-\,5400\AA) of G8 is too low to obtain good enough spectrum.
Therefore,  the lines (e.g. [\ion{O}{iii}]$\lambda5007$ and H${\beta}$) in the bluest range are excluded in the following kinematic study.

To study the kinematics of J0038+4128, we use the lines available from the obtained 1D spectra of each slice mentioned before, including emission lines of H${\alpha}$, [N~{\sc ii}]$\lambda\lambda6549,6583$, [S~{\sc ii}]$\lambda\lambda6717,6731$ and absorption lines of MgIb$\lambda\lambda5167,5172$, NaI$\lambda\lambda5889,5895$.
The emission lines represent the properties of gas ionized by AGN activity in either broad line region (BLR) or narrow line region (NLR) and the absorption lines represent the properties of stars governed by the host galaxies.
The spectra of those emission and absorption lines of different slices are presented in Figs.\, \ref{line_p1} \,\&\, \ref{line_p2}, respectively.
As Fig.\, \ref{line_p1} shows, the qualities of emission lines are high enough for fitting.
However, the fits for absorption lines is more difficult since the qualities of the observed absorption lines is relatively poor considering their intrinsic weak strengths.
Here, we only perform fits for absorption lines in the slices with signal-to-noise ratio (SNR)\footnote{The SNR of the absorption lines are defined as:  the absolute peak value of the absorption lines over the standard deviation of nearby continuum (1$\sigma$).} better than 3$\sigma$.
For some important regions (i.e. A/B and E/F slices) with absorption lines detectable smaller than 3$\sigma$, two slices are binned together to obtain good enough qualities ($\geq\,3\sigma$).
The SNRs of absorption lines of different slices are presented in Table\,2.
We fit those observed line profiles by singe/multiple Gaussian(s) to obtain their central wavelengths and full width at half-maximums (FWHMs, without subtracting instrumental broadening) .
For H${\alpha}$ emission lines of B,C,D,E,F slices, both broad and narrow line regions are detectable.
We then fit the observed line profiles by one Gaussian for [N~{\sc ii}]$\lambda$6549, one Gaussian for [N~{\sc ii}]$\lambda$6583, a pair of Gaussians for H${\alpha}$ broad and narrow components with the same central wavelength and a linear polynomial for the continuum.
The fits to other (emission and absorption) lines are similar to that of the H$\alpha$ and [N~{\sc ii}] lines.
All the fit results of both emission and absorption lines are listed in Tables\,1\&\,2, respectively.
The line-of-sight (LOS) velocities of different lines (four emission lines and three absorption lines) relative to a average redshift\footnote{The average redshift, 0.0732, is the mean measurements  of the emission lines of H${\alpha}$, [N~{\sc ii}]$\lambda 6583$, [S~{\sc ii}]$\lambda\lambda6717,6731$.} of the J0038+4128, 0.0732, as a function of different regions (slices) are shown in Fig. \ref{velocity}.

\subsubsection{Relative LOS velocity}
The relative LOS velocity is very important parameter and had been added into some of dual AGN simulation (Wang \& Yuan 2012; Blecha et al. 2013).
The J0038+4128S is represented by slices A,B,C,D,E,F  and J0038+4128N is represented by slices F,G,H,I.
The relative LOS velocity  between the J0038+4128S and J0038+4128N is about 150 km s$^{-1}$.
The result is substantially smaller than the value derived by H14, which is 453\,$\pm$\,87 km s$^{-1}$.
The discrepancy is possible because the result of H14 is biased using two sets of spectra obtained by different instruments and wavelength ranges.
The newly derived relative LOS velocity is still normal for dual AGN (generally 50-600 km\,s$^{-1}$; e.g. Comerford et al. 2012; Fu et al. 2011a, b, 2012; Barrows et al. 2013).

\subsubsection{Kinematics of emission and absorption lines}

Gaseous and stellar disks are largely co-rotated in normal galaxies,  but  counter-rotating cases of gaseous disk with respect to a stellar disk are also seen in few galaxies (Kuijken et al. 1996; Zeilinger et al. 2000; Garc{\'{\i}}a-Burillo et al. 2003; Crocker et al. 2009; Nixon et al. 2012; Johnston et al. 2013; Ricci et al. 2014).
The counter-rotating disks are possibly induced by major merger (Corsini 2014).
But the detailed mechanism of this scenario is still unclear and needs more constraints from observations.
Villforth \& Hamann (2015) perform the first try to investigate the relation between gaseous and stellar rotation disks in four double-peaked [\ion{O}{iii}]$\lambda5007$ AGN.
The J0038+4128 studied here, is essentially a dual AGN candidate, which also can provide important clues to test the (counter) rotating disk mechanisms.

With the measured emission (represent gas) and absorption (represent stellar) lines in Tables\, 1 \& 2, we compare gaseous disk in J0038+4128S with stellar disks in J0038+4128S and compare gaseous disk in J0038+4128N with stellar disks in J0038+4128N. The LOS velocity of emission lines and absorption lines display the same trend in both J0038+4128S and J0038+4128N revealing that the gaseous and stellar disks are in a co-rotating system.
As Fig. \ref{slitimages} shows, the tidal features indicate the dual AGN J0038+4128 is in early-merger stage.  The J0038+4128, combining with other three dual AGN (SDSS0952+2552, SDSS1151+4711, SDSS1502+1115) found by Villforth et al. (2015), show a co-rotating gaseous disk with respect to stellar disk. Counter-rotating between gaseous and stellar disks is however found in the only dual AGN SDSS1715+6008 (Villforth et al. 2015) which is in late-merger stage. 
The current results may imply that co-rotating between gaseous and stellar disks is a normal behaviour of dual AGN in early-merger stage and count-rotating between gaseous and stellar disks may happen in the late-merger stage of dual AGN.  However, to better understand the relations between kinematics of gaseous and stellar disks and merger stages, a larger sample of dual AGN with 2-D spectroscopy is needed.

\begin{table}[H]
\begin{center}
\caption{Properties of the five knots.}
\begin{tabular}{lccccccl}
\hline
Knots&	$\Delta D_S$ & $\Delta D_N$&	Radii&	FWHM	     &	log($\left[ \textrm{NII} \right]$/H${\alpha}$)&	 $\Delta V_k$  &	Location  	\\
	 &	(kpc)&(kpc)  &(kpc)	&(km s$^{-1}$)  &  & (km s$^{-1}$) &  \\
\hline	
$A_k$&7.15	&-- 	    &0.146	&226 &-0.21$\pm$0.02	&149 &	south edge	\\
$B_k$&5.96	&-- 	    &0.117	&315 &-0.04$\pm$0.01	&166 &	on arm	    \\
$E_k$&1.82	&3.12	&0.179	&479 &0.22$\pm$0.01	&50  &	on arm 	\\
$H_k$&--	    &2.57   &0.204	&317 &-0.28$\pm$0.03	&101 &	on arm    \\
$I_k$&--    &7.65   &0.496	&273 &-0.22$\pm$0.02	&54  &	north edge	\\

\hline
\end{tabular}
\label{knots}
\end{center}
 \begin{flushleft}

\quad \quad \quad \quad \quad  Note: $\Delta D_S$: Separation between knot and J0038+4128S.\\
\quad \quad \quad \quad \quad  $\Delta D_N$: Separation between knot and J0038+4128N.\\
\quad \quad \quad \quad \quad  $\Delta V_k$: Approximate velocity difference between knot and AGN.\\

 \end{flushleft}

\end{table}

\subsubsection{Knots}

As Fig. \ref{slitimages} shows, there are more than a dozen compact knots scattering around the two nuclei of J0038+4128. We can infer that this dual AGN has both gas rich predecessor galaxies which should be responsible for triggering this dual AGN and the surrounding knots. These knots are naturally products of galaxy merger (Villar-Mart\'in et al. 2011) and their physical properties are very important to study the merger galaxy. For the current spectroscopic observation, five knots locate in the long slit. We can extract 1D spectra of these knots within the slices labelled in Fig. \ref{slitimages}, which can provide us a opportunity to study the detail properties of these five knots.

The spectra of five knots ($A_k$,$B_k$,$E_k$,$H_k$,$I_k$) are presented in Fig.\ \ref{line_p1} and their properties of these knots are shown in the Table \ref{knots}.
There are obvious discrepancy about their FWHM between the knots on the arm and the knots on the edge, but other parameters show no discrepancies.
As shown in Table \ref{knots}, the average FWHM of H${\alpha}$ emission line of the three on arm knots is 370 km s$^{-1}$, which is larger than that 250 km s$^{-1}$ of the two knots on the edge. The knots $A_k$ and $I_k$ on the edge are disconnected in the Fig. \ref{slitimages} and the 2D-spectra Fig. \ref{slitspectra}, which can infer that the knots $A_k$ and $I_k$ are being tore apart and threw away from center.

The knots do show a wide range of its size ranging from several pc up to 400 pc (Miralles-Caballero et al. 2011). ULIRGs associated with spirals, interacting galaxies and mergers appears to proceed in larger clumps with sizes in the 0.1$-$1.5 kpc range (Elmegreen et al. 2009; F{\"o}rster Schreiber et al. 2011). Radii of the knots are correspond to the half-light radius (Whitmore et al. 1993; surace et al. 1998) but not strictly. This five knots have a average size of 0.228 kpc which is normal for the dual AGN. The scatter compacted knots around this dual AGN can testify the strong star formation activity. Merging system has extremely high star formation activity (Chapman et al. 2003; Frayer et al. 2003; Engel et al. 2010).

\section{SUMMARY}
\label{sect:SUMMARY}
We present a kinematic study of the dual AGN J0038$+$4128 based on long-slit spectroscopy obtained by LJT at Yunnan observatories. From the long-slit spectroscopy, we find that the offset velocity between J0038+4128N and J0038+4128S is about 150 km s$^{-1}$. We also study the velocity trend between emission lines (ionized by gas) and absorption lines (governed by host galaxy) and find they show the same trend.
Combing with the long-slit spectroscopy study of other four dual AGN candidates by Villforth et al. (2015), the co-rotating between gaseous and stellar disks is possibly at the early merge state and the count-rotating is possibly at the late merge state.
However, a larger sample of dual AGN with 2-D spectroscopy is still needed to better understand the relations between kinematics of gaseous and stellar disks and merger stages.

This dual AGN shows strong tidal morphologies, with more than a dozen compact knots scatter around its host galaxies. 
We have also studied the properties (e.g. FWHM, size) of five compact knots. 
These scatter compact knots infer that this dual AGN are processing strong star formation activity.

We had started a systematic search of dual AGN  since 2014 using the YFOSC on LJT of Yunnan observatories. More dual AGN and their kinematic studies will be presented in the future work.

 \section*{Acknowledgements}

We acknowledge the support of the staff of the Lijiang 2.4 m telescope. 
Fund for the telescope has been provided by CAS and the People's Government of Yunnan Province. 
We thank Fang Wang, Xu-Liang Fan, Cheng Cheng, Ju-Jia Zhang, Wei-Min Yi, Neng-Hui Liao for help. 
The work of J. M. Bai is supported by the NSFC (grants 11133006, 11361140347) and the Strategic Priority Research Program ``The Emergence of Cosmological Structures'' of the Chinese Academy of Sciences (grant No. XDB09000000).
Y. Huang and X.-W. Liu acknowledge support by the National Key Basic Research Program of China 2014CB845700.


\begin{thebibliography}{}

\bibitem[Ballo et al. (2004)]{Ballo04}
Ballo L., Braito V., Della C., R., Maraschi L., Tavecchio F., Dadina M., 2004, \apj, 600, 634

\bibitem[Barrows et al.(2012)]{2012ApJ...744....7B} Barrows, R.~S., Stern,
D., Madsen, K., et al.\ 2012, \apj, 744, 7

\bibitem[Barrows et al.(2013)]{2013ApJ...769...95B} Barrows, R.~S.,
Sandberg Lacy, C.~H., Kennefick, J., et al.\ 2013, \apj, 769, 95


\bibitem[Begelman et al. (1980)]{Begelman80}
Begelman M. C., Blandford R. D., Rees M. J., 1980, \nature, 287, 307

\bibitem[Bianchi et al. (2008)]{Bianchi08}
Bianchi S., Chiaberge M., Piconcelli E., Guainazzi M., Matt G., 2008, \mnras, 386, 105


\bibitem[Blecha et al.(2013)]{2013MNRAS.429.2594B} Blecha, L., Loeb, A.,
\& Narayan, R.\ 2013, \mnras, 429, 2594

\bibitem[{{Chapman} {et~al.}(2003){Chapman}, {Windhorst}, {Odewahn}, {Yan}, \&
  {Conselice}}]{Chapman03}
{Chapman}, S.~C., {Windhorst}, R., {Odewahn}, S., {Yan}, H., \& {Conselice}, C.
  2003, \apj, 599, 92

\bibitem[Colpi and Dotti (2011)]{Colpi and Dotti11}
Colpi M., Dotti M., 2011, ASdv. Sci. Lett., 4, 181


\bibitem[Comerford et al.(2009)]{2009ApJ...698..956C} Comerford, J.~M.,
Gerke, B.~F., Newman, J.~A., et al.\ 2009, \apj, 698, 956

\bibitem[Comerford et al.(2009)]{2009ApJ...702L..82C} Comerford, J.~M.,
Griffith, R.~L., Gerke, B.~F., et al.\ 2009, \apjl, 702, L82


\bibitem[Comerford et al.(2011)]{2011ApJ...737L..19C} Comerford, J.~M.,
Pooley, D., Gerke, B.~F., \& Madejski, G.~M.\ 2011, \apjl, 737, LL19

\bibitem[Comerford et al.(2012)]{2012ApJ...753...42C} Comerford, J.~M.,
Gerke, B.~F., Stern, D., et al.\ 2012, \apj, 753, 42

\bibitem[Corsini(2014)]{2014ASPC..486...51C} Corsini, E.~M.\ 2014,
Multi-Spin Galaxies, ASP Conference Series, 486, 51

\bibitem[Crenshaw et al. (1999)]{Crenshaw99}
Crenshaw D. M., Kraemer S. B., Boggess A., Maran S. P., Mushotzky R. F., Wu C. -C., 1999, apj, 516, 750

\bibitem[Crocker et al.(2009)]{2009MNRAS.393.1255C} Crocker, A.~F., Jeong,
H., Komugi, S., et al.\ 2009, \mnras, 393, 1255


\bibitem[{{Elmegreen} {et~al.}(2009){Elmegreen}, {Elmegreen}, {Marcus},
  {Shahinyan}, {Yau}, \& {Petersen}}]{Elmegreen09}
{Elmegreen}, D.~M., {Elmegreen}, B.~G., {Marcus}, M.~T., {Shahinyan}, K.,
  {Yau}, A., \& {Petersen}, M. 2009, \apj, 701, 306

\bibitem[{{Engel} {et~al.}(2010){Engel}, {Tacconi}, {Davies}, {Neri}, {Smail},
  {Chapman}, {Genzel}, {Cox}, {Greve}, {Ivison}, {Blain}, {Bertoldi}, \&
  {Omont}}]{Engel10}
{Engel}, H., {et~al.} 2010, \apj, 724, 233

\bibitem[Dunn et al. (2007)]{Dunn07}
Dunn J. P., Crenshaw D. M., Kraemer S. B., Gabel J. R., 2007, apj, 134, 1061

\bibitem[F{\"o}rster Schreiber et al.(2011)]{2011ApJ...739...45F}
F{\"o}rster Schreiber, N.~M., Shapley, A.~E., Genzel, R., et al.\ 2011,
\apj, 739, 45

\bibitem[{{Frayer} {et~al.}(2003){Frayer}, {Armus}, {Scoville}, {Blain},
  {Reddy}, {Ivison}, \& {Smail}}]{Frayer03}
{Frayer}, D.~T., {Armus}, L., {Scoville}, N.~Z., {Blain}, A.~W., {Reddy},
  N.~A., {Ivison}, R.~J., \& {Smail}, I. 2003, \aj, 126, 73


\bibitem[Fu et al.(2011a)]{2011ApJ...740L..44F} Fu, H., Zhang, Z.-Y., Assef,
R.~J., et al.\ 2011a, \apjl, 740, L44

\bibitem[Fu et al.(2011b)]{2011ApJ...733..103F} Fu, H., Myers, A.~D.,
Djorgovski, S.~G., \& Yan, L.\ 2011b, \apj, 733, 103


\bibitem[Fu et al.(2012)]{2012ApJ...745...67F} Fu, H., Yan, L., Myers,
A.~D., et al.\ 2012, \apj, 745, 67

\bibitem[Garc{\'{\i}}a-Burillo et
al.(2003)]{2003A&A...407..485G} Garc{\'{\i}}a-Burillo, S., Combes, F., Hunt, L.~K., et al.\ 2003, \aap, 407, 485

\bibitem[Hernquist (1989)]{Hernquist89}
Hernquist L., 1989, \nature, 340, 687

\bibitem[Hopkins et al. (2008)]{Hopkins08}
Hopkins P. F., Hernquist L., Cox T. J., Kere\u{s} D., 2008, \apjs, 175, 356

\bibitem[Huang et al.(2014)]{2014MNRAS.439.2927H} Huang, Y., Liu, X.-W.,
Yuan, H.-B., et al.\ 2014, \mnras, 439, 2927

\bibitem[Huerta \& Gonz{\'a}lez(2009)]{2009RMxAC..35..227H} Huerta, E.~M., \& Gonz{\'a}lez, J.~J.\ 2009, Revista Mexicana de Astronomia y Astrofisica Conference Series, 35, 227


\bibitem[Junkkarinen et al. (2001)]{Junkkarinen01}
Junkkarinen V., Shields G. A., Beaver E. A., Burbidge E. M., Cohen R. D., Hamann F., Lyons R. W., 2001, \apj, 549, 155

\bibitem[Johnston et al.(2013)]{2013MNRAS.428.1296J} Johnston, E.~J.,
Merrifield, M.~R., Arag{\'o}n-Salamanca, A.,
\& Cappellari, M.\ 2013, \mnras, 428, 1296

\bibitem[Kauffmann and Haehnelt (2000)]{Kauffmann and Haehnelt00}
Kauffmann G., Haehnelt M. 2000, \mnras, 311, 576

\bibitem[Keel(1996)]{1996ApJS..106...27K}
 Keel, W.~C.\ 1996, \apjs, 106, 27

\bibitem[Komossa et al. (2003)]{Komossa03}
Komossa S., Burwitz V., Hasinger G., Predehl P., Kaastra J. S., Ikebe Y., 2003, \apj, 582L, 15


\bibitem[Kormendy
\& Ho(2013)]{2013ARA&A..51..511K} Kormendy, J., \& Ho, L.~C.\ 2013, \araa, 51, 511

\bibitem[Koss et al. (2011)]{Koss11}
Koss M., et al., 2011, \apj, 735L, 42

\bibitem[Kuijken et al.(1996)]{1996MNRAS.283..543K} Kuijken, K., Fisher,
D., \& Merrifield, M.~R.\ 1996, \mnras, 283, 543


\bibitem[Liu et al. (2010)]{Liu10}
Liu X., Greene J. E., Shen Y., Strauss M. A., 2010, \apj, 715L, 30

\bibitem[Liu et al. (2013)]{Liu13}
Liu X., Civano F., Shen Y., Green P., Greene J. E., Strauss M. A., 2013, \apj, 762, 110

\bibitem[{\L}okas et al.(2014)]{2014MNRAS.445.1339L} {\L}okas, E.~L.,Athanassoula, E., Debattista, V.~P., et al.\ 2014, \mnras, 445, 1339

\bibitem[McGurk et al. (2011)]{MuGurk11}
McGurk R. C., Max C. E., Rosario D. J., Shields G. A., Smith K. L., Wright S. A., 2011, \apj, 738, L2
\bibitem[McGurk et al.(2012)]{2012AAS...21922505M} McGurk, R.~C., Max,
C.~E., Rosario, D.~J., et al.\ 2012, American Astronomical Society Meeting
Abstracts \#219, 219, \#225.05

\bibitem[McGurk et al.(2014)]{2014IAUS..304..371M} McGurk, R.~C., Max,
C.~E., Medling, A., \& Shields, G.~A.\ 2014, IAU Symposium, 304, 371

\bibitem[Milosavljevic and David (2001)]{Milosavljevic and David01}
Milosavljevi\'c M., Merritt, D., 2001, \apj, 563, 34

\bibitem[Miralles-Caballero et al.(2011)]{2011AJ....142...79M}
Miralles-Caballero, D., Colina, L., Arribas, S.,
\& Duc, P.-A.\ 2011, \aj, 142, 79

\bibitem[Miwa \& Noguchi(1998)]{1998ApJ...499..149M} Miwa, T., \& Noguchi, M.\ 1998, \apj, 499, 149


\bibitem[Noguchi(1987)]{1987MNRAS.228..635N} Noguchi, M.\ 1987, \mnras,228, 635

\bibitem[Nixon et al.(2012)]{2012MNRAS.422.2547N} Nixon, C.~J., King,
A.~R., \& Price, D.~J.\ 2012, \mnras, 422, 2547

\bibitem[Ricci et al.(2014)]{2014MNRAS.440.2419R} Ricci, T.~V., Steiner,
J.~E., \& Menezes, R.~B.\ 2014, \mnras, 440, 2419

\bibitem[Richstone et al.(1998)]{1998Natur.395A..14R} Richstone, D., Ajhar, E.~A., Bender, R., et al.\ 1998, nat, 395, A14

\bibitem[Rosario et al. (2011)]{Rosario11}
Rosario D. J., McGurk R. C., Max C. E., Shields G. A., Smith K. L., Ammons S. M., 2011, \apj, 739, 44

\bibitem[Salo(1991)]{1991A&A...243..118S} Salo, H.\ 1991, \aap, 243, 118

\bibitem[Shields et al. (2012)]{Shields12}
Shields G. A., Rosario D. J., Junkkarinen V., Chapman C., Bonning E. W., Chiba T., 2012, \apj, 744,151

\bibitem[{{Surace} {et~al.}(1998){Surace}, {Sanders}, {Vacca}, {Veilleux}, \&
  {Mazzarella}}]{Surace98}
{Surace}, J.~A., {Sanders}, D.~B., {Vacca}, W.~D., {Veilleux}, S., \&
  {Mazzarella}, J.~M. 1998, \apj, 492, 116

\bibitem[Villforth \& Hamann(2015)]{2015arXiv150100325V} Villforth, C., \& Hamann, F.\ 2015, arXiv:1501.00325

\bibitem[Villar-Martin et al. (2011)]{Villar-Martin 11}
Villar-Mart\'in M., Humphrey A., Delgado R. G., Colina L., Arribas S., 2011, \mnras, 418, 2032

\bibitem[Villforth \& Hamann(2015)]{2015AJ....149...92V} Villforth, C., \& Hamann, F.\ 2015, \aj, 149, 92

\bibitem[Wang \& Yuan(2012)]{2012MNRAS.427L...1W} Wang, X.-W., \& Yuan, Y.-F.\ 2012, \mnras, 427, L1

\bibitem[{{Whitmore} {et~al.}(1993){Whitmore}, {Schweizer}, {Leitherer},
  {Borne}, \& {Robert}}]{Whitmore93}
{Whitmore}, B.~C., {Schweizer}, F., {Leitherer}, C., {Borne}, K., \& {Robert},
  C. 1993, \aj, 106, 1354

\bibitem[Yu et al. (2011)]{Yu11}
Yu Q. -J., Lu Y. -J., Mohayaee R., Colin, J., 2011, \apj, 738, 92

\bibitem[Zeilinger et al.(2000)]{2000ASPC..215..214Z} Zeilinger, W.~W.,
Vega Beltr{\'a}n, J.~C., Rozas, M., et al.\ 2000, Cosmic Evolution and
Galaxy Formation: Structure, Interactions, and Feedback, 215, 214

\bibitem[Zhang et al. (2012)]{zhang12}
Zhang J. J., Fan Y. F., Chang L., Wang C. J., \& Yi W. M, 2012, Astronomical Research \& Technology (Chinese), 9, 411

\end{thebibliography}
\end{document}